\documentclass[aps,prl,twocolumn,amssymb,amsmath]{revtex4}

\usepackage{graphicx}
\usepackage{subfigure}
\usepackage{lmodern}
\usepackage{textcomp}
\usepackage[T1]{fontenc}

\begin{document}

\title{Experimental demonstration of interaction-free all-optical switching via the quantum Zeno effect}

\author{Kevin T. McCusker}
\email{kevin.mccusker@northwestern.edu}
\author{Yu-Ping Huang}
\author{Abijith Kowligy}
\author{Prem Kumar}
\affiliation{Center for Photonic Communication and Computing, EECS Department, Northwestern University, 2145 Sheridan Road, Evanston, Illinois 60208-3118, USA}

\date{\today}

\begin{abstract}
We experimentally demonstrate all-optical interaction-free switching using the quantum Zeno effect, achieving a high contrast of 35:1. The experimental data matches a zero-parameter theoretical model for several different regimes of operation, indicating a good understanding of the switch's characteristics. We also discuss extensions of this work that will allow for significantly improved performance, and the integration of this technology onto chip-scale devices.
\end{abstract}

\maketitle

Interaction-free measurement \cite{interactionfree_proposal1, interactionfree_proposal2, interactionfree_proposal3, interactionfree_experiment1} allows observation to be made in a regime that is impossible classically, i.e., without the interaction actually occurring. Incorporating the quantum Zeno effect, such measurement can even be done with arbitrarily high efficiency \cite{ high_efficiency_zeno}, allowing for exotic experiments such as counterfactual quantum computation \cite{counterfactual}. In this Letter, we describe implementation of a recently-proposed protocol \cite{coherent_zeno_switch, incoherent_and_coherent_zeno_switch} and for the first time demonstrate high-contrast all-optical switching based on the quantum Zeno effect.

All-optical switching will allow for fast, efficient networks with ultrahigh data capacity \cite{optical_transistors}. However, common switches involving nonlinear optical devices employ direct coupling between the signal and the pump, which causes photon loss and possibly, in a quantum system, decoherence as well. In contrast, a Zeno-based switch mediated via an interaction-free process has no direct coupling, so the loss is minimized and the decoherence from signal-pump coupling is eliminated, making it suitable for working with quantum as well as classical signals. In addition, by using a microresonator instead of bulk optics (see discussion at the end of this Letter), such a switch can operate with ultra-low energies (potentially down to the single-photon level) and can even act as an optical transistor, i.e., a lower-energy pulse can switch a higher-energy pulse.

\begin{figure}
    \centering
    \subfigure
    {
        \includegraphics[width=3.3in]{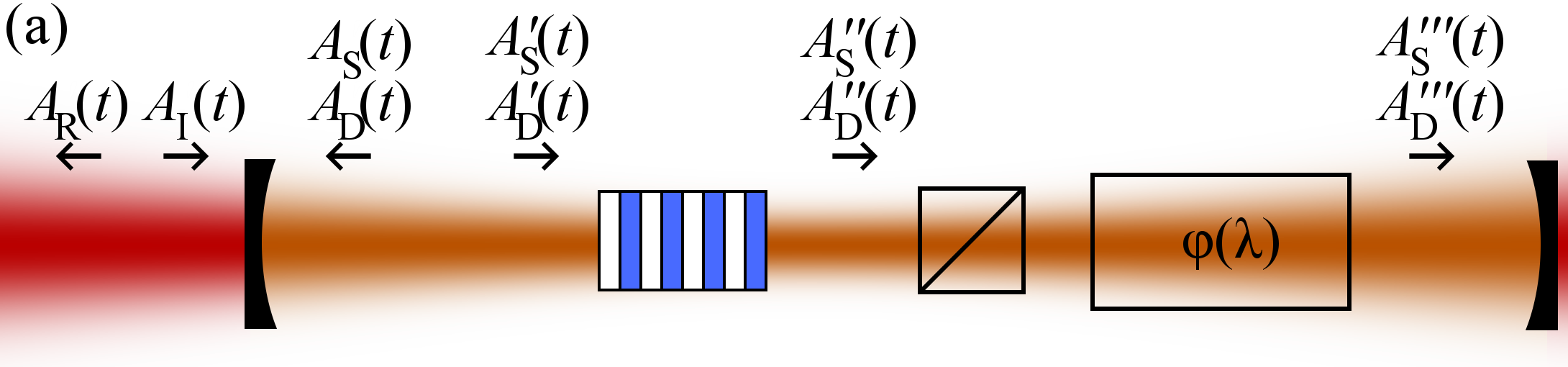}
        \label{fig:switch_graphic}
    }
    \subfigure
    {
        \includegraphics[width=3.3in]{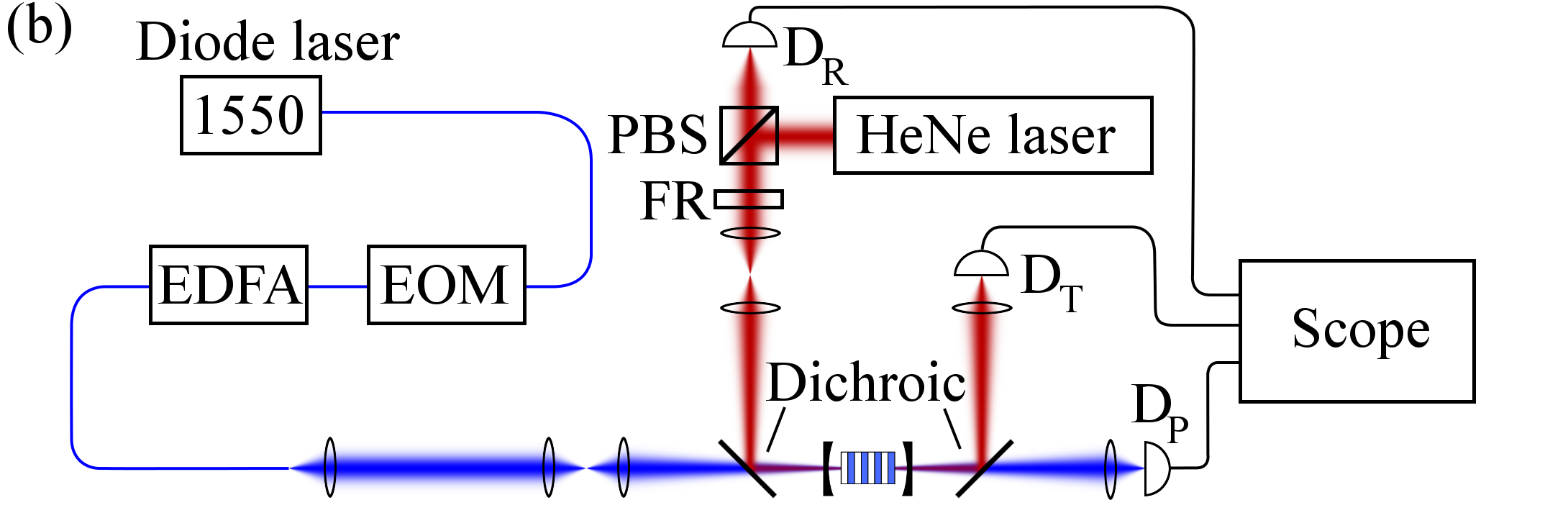}
        \label{fig:experimental_setup}
    }
    \caption{\subref{fig:switch_graphic} The electric fields at various points throughout the cavity (see text for details). \subref{fig:experimental_setup} The experimental setup. EDFA: Erbium-doped fiber amplifier. EOM: Electro-optic modulator. PBS: polarizing beam-splitter. FR: Faraday rotator (so the light reflected from the cavity is transmitted through the PBS instead of reflected). D: detectors for pump, transmitted signal, and reflected signal.}
    \label{fig:graphic_and_setup}
\end{figure}

The prototype switch presented in this letter is based on a Fabry-P\'{e}rot design (see Fig.~\ref{fig:switch_graphic}) with an intracavity crystal phase-matched for difference-frequency (DF) generation (other interactions such as sum-frequency generation would work as well). The cavity is resonant with a high finesse at both the signal and the difference frequencies (but not at the pump frequency for this implementation; a high finesse for the pump would decrease the required pump power). In the normal operation of the Fabry-P\'{e}rot, i.e., with the pump off, when a signal photon (or pulse) reaches the cavity, a small portion of its amplitude initially enters the cavity, and, upon successive round-trips, constructively interferes with the incoming amplitude, allowing the entire photon to pass through cavity, with only a small overall reflection. With the pump on, however, the signal field that enters the cavity is converted to the DF field, so constructive interference is inhibited, and the photon is prevented from entering the cavity. From the Zeno perspective, the pump is constantly measuring if the photon is in the cavity, which guarantees that the photon will \textit{not} enter the cavity (or even ever interact with the pump!), and instead will be reflected by the cavity.

Besides our $\chi^{(2)}$-based implementation, other protocols for Zeno-based all-optical switching have also been proposed, employing, e.g., cavity-enhanced two-photon absorption (TPA) by rubidium atomic vapor \cite{zeno_tpa_proposal} or inverse-Raman scattering in silicon-based microresonators \cite{zeno_modulation_theory_gaeta} (which has been demonstrated for modulation only, not switching \cite{zeno_modulation_gaeta}). Initial evidence of the TPA-induced switching has been recently observed with very low contrast, where the signal transmission through the switch was shown to be affected by about $3\%$ \cite{Hendrickson:12, zeno_tpa_franson}.

The implementation presented in this Letter is a modified version of the proposal developed in \cite{incoherent_and_coherent_zeno_switch}. The primary difference is that we use a continuous-wave (CW) signal beam for experimental convenience (the pump is still pulsed). The switch is modeled here using quasi-static analysis, similar to the usual classical description of a  Fabry-P\'{e}rot cavity, which is valid if all of the input intensities vary slowly with respect to the cavity round-trip time. Let the fields of the signal and the DF in the cavity be denoted by $A_\text{S}$ and $A_\text{D}$, respectively. The various fields at the first mirror are related via (see Fig.~\ref{fig:switch_graphic} for a pictorial representation of the location of these fields) $A_\text{S}'(t)=A_\text{S}(t)r_\text{S}+A_\text{I}(t)t_\text{S}$, $A_\text{R}(t)=A_\text{S}(t)t_\text{S}-A_\text{I}(t)r_\text{S}$, and $A_\text{D}'(t)=A_\text{D}(t)r_\text{D}$, where $r_\text{S}$ ($r_\text{D}$) and $t_\text{S}$ ($t_\text{D}$) refer to the reflection and transmission coefficients, respectively, at the signal (difference) frequency, and $A_\text{I}(t)$ and $A_\text{R}(t)$ are the signal fields at time $t$ for the incident light and the reflected light, respectively (for the DF, only the field inside of the cavity is considered). After the first mirror, the fields undergo three-wave mixing in the nonlinear crystal, which---assuming a single-mode regime, perfect phase-matching, and an undepleted pump---gives:
\begin{align*}
A_\text{S}''(t)=A_\text{S}'(t)&\cos(g\sqrt{I_\text{P}})
+(\frac{\omega_\text{S}}{\omega_\text{D}})^{1/2}A_\text{D}'(t)\sin(g\sqrt{I_\text{P}}), \\
A_\text{D}''(t)=A_\text{D}'(t)&\cos(g\sqrt{I_\text{P}})
-(\frac{\omega_\text{D}}{\omega_\text{S}})^{1/2}A_\text{S}'(t)\sin(g\sqrt{I_\text{P}}). 
\end{align*}
Here, $\omega_\text{S}$ ($\omega_\text{D}$) is the signal (difference) frequency and $g\sqrt{I_\text{P}}$ is the strength of the interaction, which depends on the nonlinear coefficient of the crystal, the focusing conditions, the crystal length, and the time-dependent pump power $I_\text{P}$. After the crystal, there is some loss from, e.g., scattering or absorption in the crystal or mirrors, followed by a frequency-dependent phase shift, which takes into account the variable optical path-length of the cavity, giving $A_\text{S}'''(t)=A_\text{S}''(t)\eta_\text{S}e^{i\phi_\text{S}}$ and $A_\text{D}'''(t)=A_\text{D}''(t)\eta_\text{D}e^{i\phi_\text{D}}$. The fields then reach the second mirror, where we can determine the output of the cavity by applying similar transformations as the first mirror. After propagating back through the cavity following the same loss and phase transformations, the fields return to the first mirror for the start of the next round-trip, completing the cycle, and yielding $A_\text{S}(t+\Delta t)=A_\text{S}'''(t)r_\text{S}\eta_\text{S}e^{i\phi_\text{S}}$ and $A_\text{D}(t+\Delta t)=A_\text{D}'''(t)r_\text{D}\eta_\text{D}e^{i\phi_\text{D}}$,
where $\Delta t$ is the cavity round-trip time. The parameters in these equations can be directly measured, leaving no free parameters for describing the experimental performance of the switch.

Our experimental setup is shown in Fig.~\ref{fig:experimental_setup}. We use the output from a CW Helium-Neon laser at 633\,nm as the signal with a 1550-nm pulse created by chopping the output of a CW laser with an electro-optic modulator and then amplifying it with an erbium-doped fiber amplifier. The cavity is made up of two curved mirrors (radii of curvature of 75\,mm, separated by about 25\,mm) around a 5-mm-long lithium-niobate crystal, which is periodically poled for quasi-phase-matched DF generation at 1070\,nm.  The position of one mirror can be scanned with a piezo-electric actuator, giving us one of the two degrees of freedom necessary to tune the cavity resonance for two different frequencies. For the other, we can adjust the temperature of the crystal, which changes the frequency-dependent path length (the phase matching is also affected, but not significantly).

\begin{figure}
    \centering
    \includegraphics[width=3.4in]{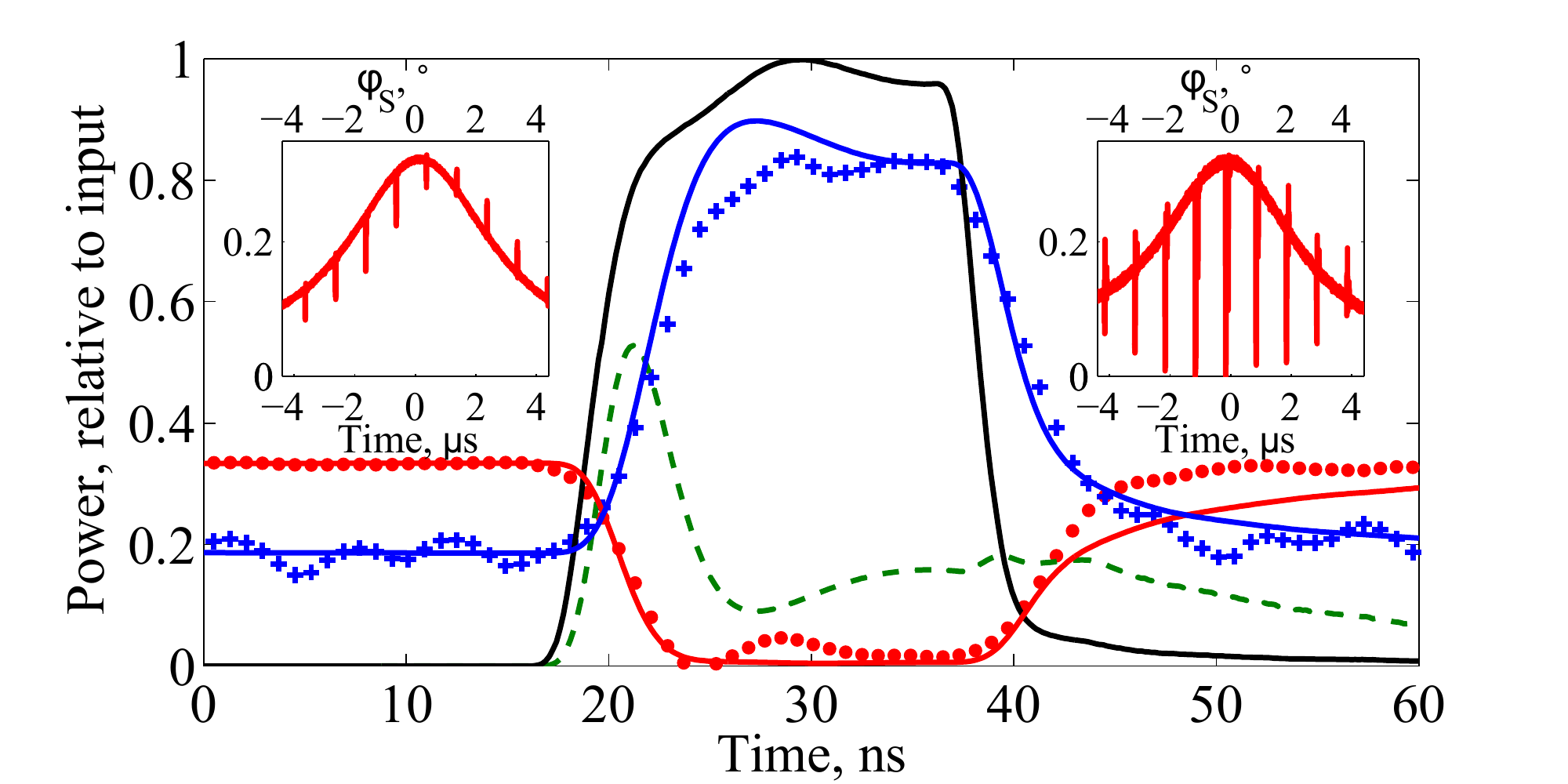}
    \caption{Experimental switching performance when both the signal and DF fields are on resonance ($\phi_\text{S}=\phi_\text{D}=0$). Red dots: transmitted signal (measured). Red line: transmitted signal (theory). Blue pluses: reflected signal (measured). Blue line: reflected signal (theory). Black line: pump pulse (measured and then used to predict the transmitted and reflected signals). Green dashed line: (theoretical) signal lost to DF generation. Note the theory curves are not fits, but zero-parameter predictions. All powers (except for the pump) are to scale relative to the input signal power. Pump pulse width is 20\,ns with a peak power of 17\,W. Insets: Transmitted signal as the piezo-mounted mirror is scanned while the pump pulses are periodically applied. When the scan is asymmetric (left), $\phi_\text{D} \neq 0$, and when it is symmetric (right), $\phi_\text{D}=0$.}
    \label{fig:experiment_switching}
\end{figure}

In order to predict the switching behavior, we first need to determine the experimental values for the parameters in the above equations. The values of the cavity finesse (ratio of the free-spectral range to the bandwidth) are 28.3 and 276 at the signal and difference frequencies, respectively, and the mirror reflectivity is measured to be 0.938 at the signal frequency. From this, we can determine $r_\text{S}=0.968$, $t_\text{S}=0.250$, and $\eta_\text{S}=0.977$. Since it does not matter where in the cavity the DF is lost, we need to only determine the overall transmission coefficient for this frequency for one round-trip through the cavity, which is found to be $r_\text{D}^2\eta_\text{D}^2=0.989$.  The single-pass depletion of the signal is measured to be 0.65$\%$ at a peak pump power of 13\,W, corresponding to $g=0.022/\sqrt{\text{W}}$. Measuring the light transmitted through the cavity with the pump off as we scan the piezo-mounted mirror allows us to determine $\phi_\text{S}$, but finding $\phi_\text{D}$ is more difficult. To determine this parameter, we send in pump pulses every \textmu s while scanning the cavity mirror at a speed to traverse the cavity bandwidth in about 20\,\textmu s. This allows us to see the switching behavior for several different values of $\phi_\text{S}$ and $\phi_\text{D}$. The values of $\phi_\text{S}$ can be directly measured, whereas the values of $\phi_\text{D}$ are revealed relative to each other up to a single absolute offset, since no other phase shifts from sources such as mechanical vibrations or thermal drift occur on the \textmu s time scale. The switching behavior (both theoretically and experimentally) is asymmetric for this scan around $\phi_\text{S}=0$ unless $\phi_\text{D}$ is also 0 at the same mirror position (see Fig.~\ref{fig:experiment_switching} insets). Tuning the crystal temperature until the switching behavior becomes symmetric during this scan allows us to set $\phi_\text{D}=0$.

Typical switching performance is shown in Fig.~\ref{fig:experiment_switching}, along with the theoretical predictions, which are based on measurements discussed above with no free parameters for the switching behavior itself. We start out at $t=0$ with the pump off, so most of the signal light is transmitted by the cavity. There is still some reflected light since the signal is under coupled to the cavity owing to intracavity loss (this could be compensated for by using a lower-reflectivity first mirror). The loss in the cavity is also why we observe that $T+R\neq1$. As the pump turns on, some of the signal light is converted into the DF field, which changes the cavity conditions so as to inhibit the signal light from entering the cavity, causing the transmission to fall and the reflection to increase. As one can see from the plots in Fig.~\ref{fig:experiment_switching}, the theory and experiment agree well. The ratio between the transmitted power when the pump is off to that when it is on is 35, showing high-contrast operation of our switch.

\begin{figure}
    \centering
    \includegraphics[width=3.4in]{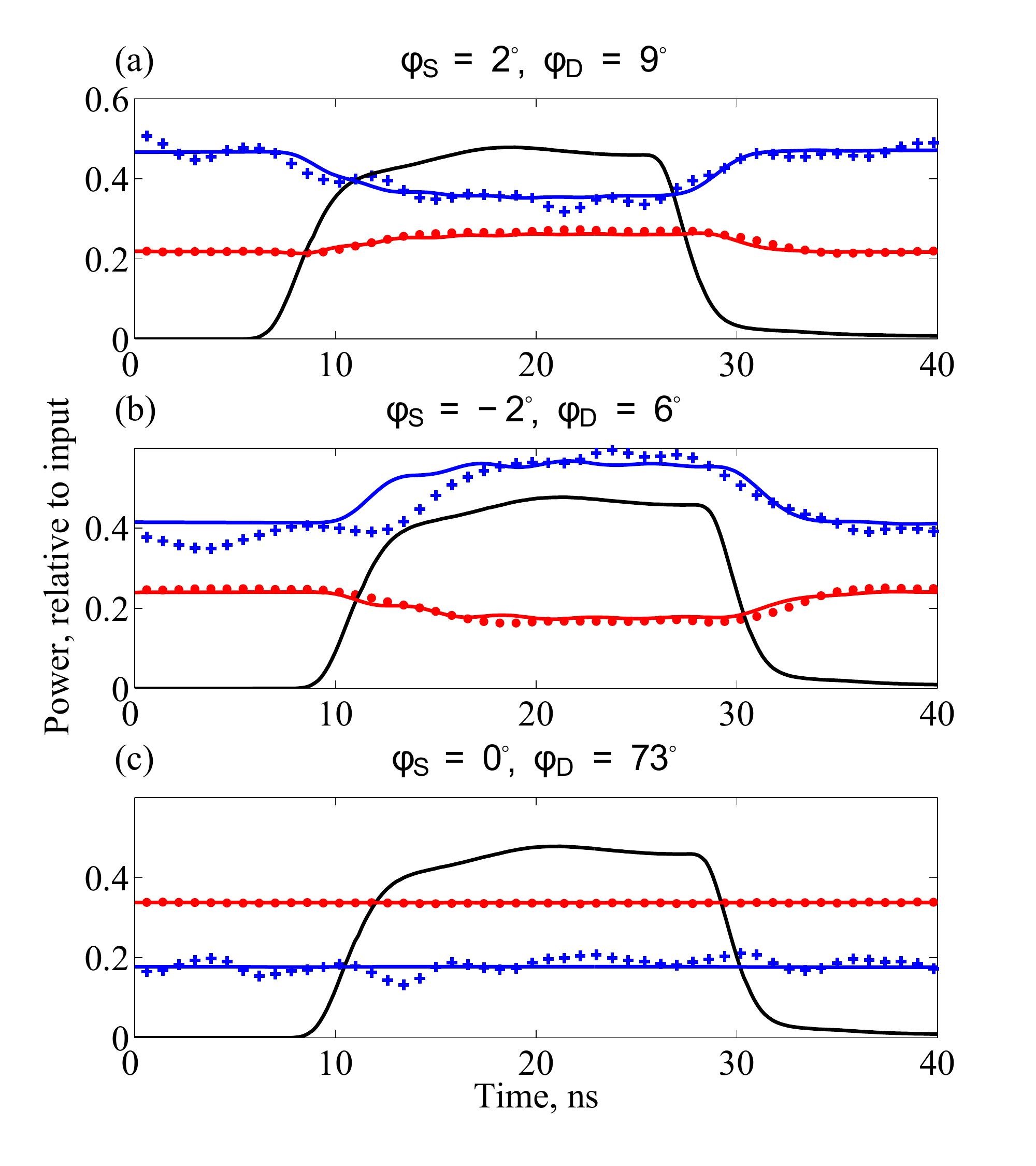}
    \caption{Switching behavior when off resonance.  When slightly off-resonance at the signal frequency, the pump can effectively shift the cavity resonance, either increasing (a) or decreasing (b) the cavity transmission, depending on the phase shift of the DF. When significantly off resonance at the DF (c), there is effectively no change in the cavity behavior when the pump is turned on. All theory curves have no fitting parameters. Pump pulse width and peak power are the same as in Fig.~\ref{fig:experiment_switching} (20~ns and 17~W, respectively).}
    \label{fig:resonance_change}
\end{figure}

In addition to studying the switch performance with both the signal and the difference frequencies on resonance, we explored off-resonant conditions. When the cavity is doubly resonant, theory predicts that the transmission of the cavity is not just lowered at the signal frequency, but in fact it is shifted to a different frequency. We can observe this shift by slightly detuning the cavity. For determining $\phi_\text{D}$ when it is $\neq0$, we observe the signal output while moving the mirror position by several multiples of the free-spectral range of the cavity at the signal frequency. Since we know at what position $\phi_\text{D}=0$, and that $\Delta{L}=\lambda_\text{S}\frac{\Delta\phi_\text{S}}{2\pi}=\lambda_\text{D}\frac{\Delta\phi_\text{D}}{2\pi}$, the value of $\phi_\text{D}$ can be determined at any position near the signal resonance (because $\phi_\text{S}$ can always be directly measured when near resonance). In Fig.~\ref{fig:resonance_change}, we show evidence of such resonance shifting. We slightly detune the cavity at the signal frequency, equivalent to the signal being slightly off-resonant with the cavity. When the DF detuning is in the same direction as the signal, the resonance is shifted towards the signal, allowing more of the light to enter and be transmitted through the cavity (Fig.~\ref{fig:resonance_change}(a)). Conversely, when the DF is shifted in the opposite direction as the signal, the resonance is shifted further away from the signal, allowing less of the light to pass through (Fig.~\ref{fig:resonance_change}(b)). If we significantly detune the cavity at the DF, then almost no switching is observed, as can be seen in Fig.~\ref{fig:resonance_change}(c). This is due to the multiple passes of DF generation destructively interfering with each other, allowing for very little overall frequency conversion.

\begin{figure}
    \centering
    \includegraphics[width=3.4in]{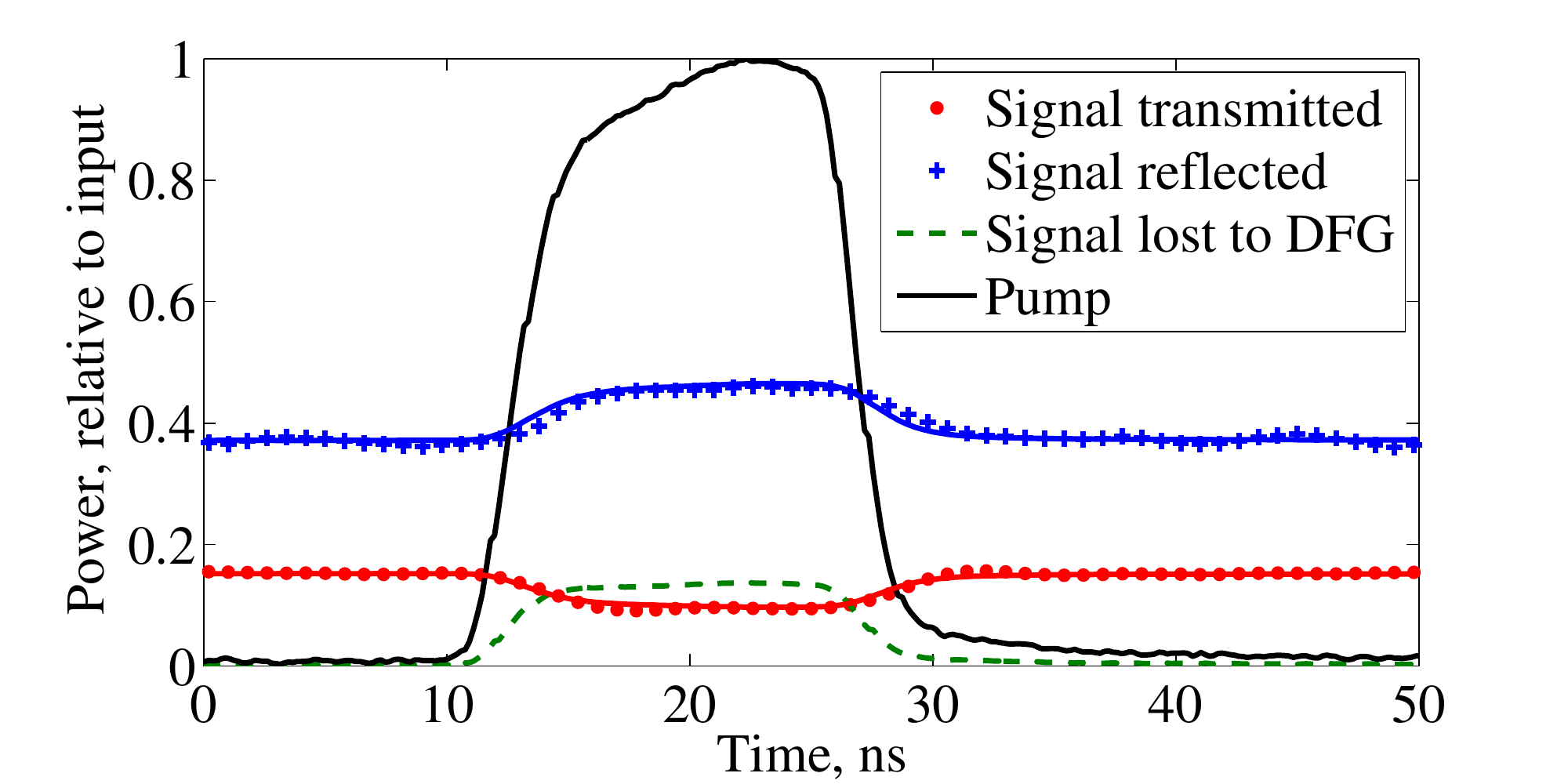}
    \caption{Switching behavior when the cavity is very lossy at the DF. Even with a significantly higher peak pump power (150\,W compared to 17\,W in Fig.~\ref{fig:experiment_switching}), the switching contrast is significantly lower (1.6:1 compared to 35:1). In this plot the pump pulse duration is 15\,ns. Note the peak transmission of the signal through the cavity is lower due to increased intracavity loss from the filter.}
    \label{fig:incoherent}
\end{figure}

The fact that the cavity is low loss at the DF has a significant effect on the performance of the switch. If the cavity is instead very lossy at this frequency, the theory predicts that much more pump power is required for switching, and the cavity resonance at the signal frequency is destroyed (rather than shifted) \cite{incoherent_and_coherent_zeno_switch}. To investigate this regime of operation, we insert a filter (Semrock, model FF01-640/14-25, transmission $<10^{-4}$ at 1070\,nm) into the cavity to reflect the DF light out. With this filter in the cavity, $\eta_\text{S}$ drops to $0.951$, and $\eta_\text{D} \approx 10^{-2}$ (when $\eta_\text{D} \ll r_\text{S}\eta_\text{S}$, its precise value is not relevant). The observed switching contrast is much worse at 1.6:1 (see Fig.~\ref{fig:incoherent}), as expected, even at the much higher peak pump power of 150\,W compared to 17\,W in Fig.~\ref{fig:experiment_switching}. While the absolute value of $\phi_\text{D}$ is difficult to measure in this case, nevertheless we are able to look at different relative values of $\phi_\text{D}$ and verify that the switching contrast does not change, as predicted.

\begin{figure}
    \centering
    \includegraphics[width=3.4in]{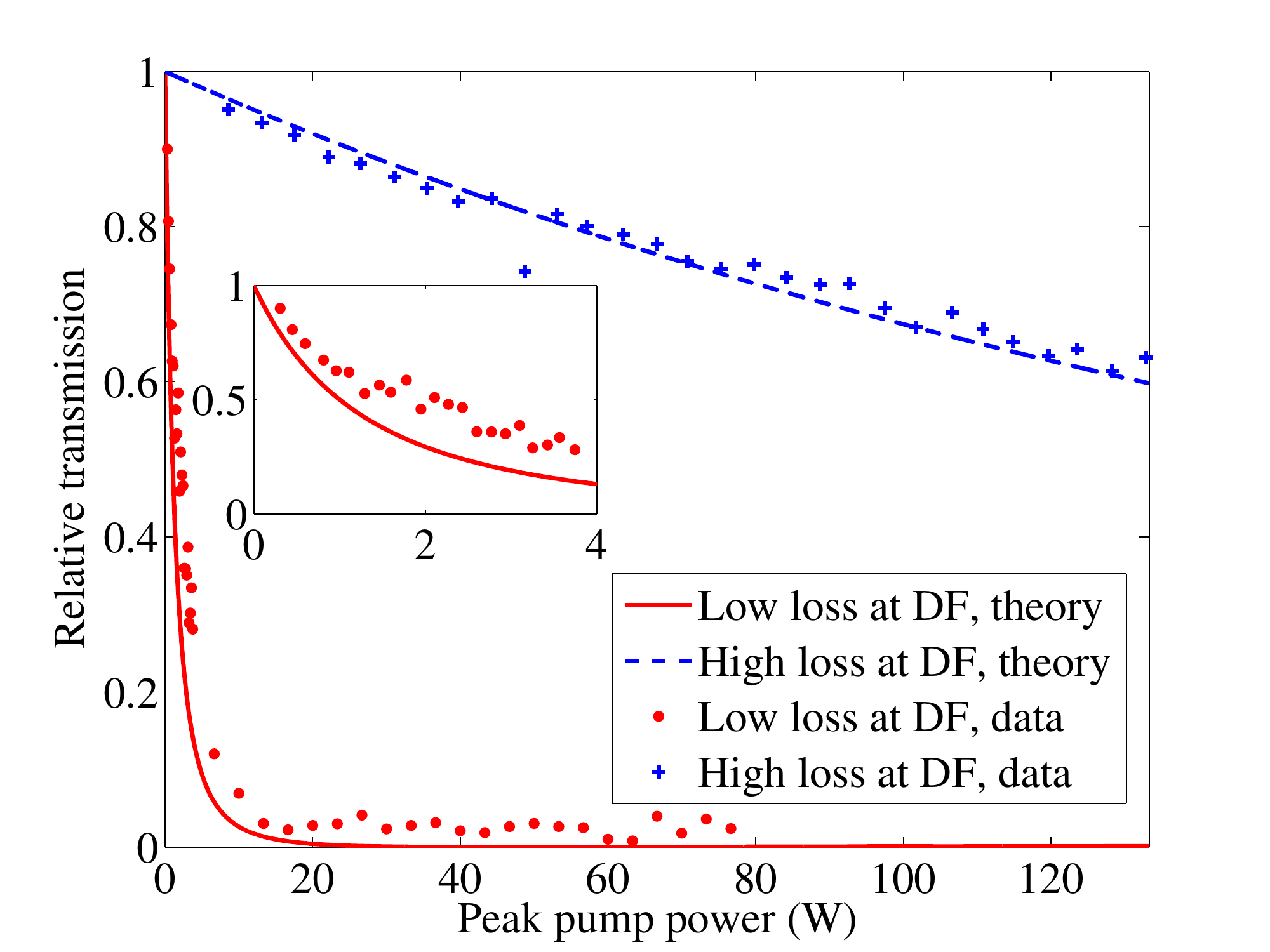}
    \caption{Relative transmission of the signal light vs. pump power. The switching does not depend on pump power beyond a certain threshold, making the device less sensitive to pump fluctuations. When the cavity is lossy at the DF, the pump power required for switching is much higher. Inset: A closer look at the switching behavior for low pump powers.}
    \label{fig:power_vs_depletion}
\end{figure}

One feature of this device is that the switching is not very dependent on the pump power. Taking the pump power to be constant in time and letting $g'=g\sqrt{I_\text{P}}$, we can solve for the steady-state transmission and reflection coefficients of the cavity when on double resonance, i.e., $\phi_\text{S}=\phi_\text{D}=0$:
\begin{align*}
t_\text{cavity}&=\frac{t^2_\text{S}\eta_\text{S}(\cos g'-r_\text{D}^2\eta_\text{D}^2)}{1-r_\text{S}^2\eta^2_\text{S}\cos g'-r_\text{D}^2\eta^2_\text{D}\cos g'+r^2_\text{S}r^2_\text{D}\eta^2_\text{S}\eta^2_\text{D}}, \\
r_\text{cavity}&=\frac{-r_\text{S}(1-\eta^2_\text{S}\cos g'-r_\text{D}^2\eta^2_\text{D}\cos g'+r^2_\text{D}\eta^2_\text{S}\eta^2_\text{D})}{1-r_\text{S}^2\eta^2_\text{S}\cos g'-r_\text{D}^2\eta^2_\text{D}\cos g'+r^2_\text{S}r^2_\text{D}\eta^2_\text{S}\eta^2_\text{D}}.
\end{align*}
The experimental measurement and theoretical prediction for the relative power of the transmitted light are shown in Fig.~\ref{fig:power_vs_depletion}. As the pump power increases, the cavity rapidly becomes highly reflective, and then saturates (until $g'$ approaches $2\pi$, at a pump power of about 80\,kW). This is clearly demonstrated in the figure for the doubly-resonant case. As predicted, significantly more pump power is required to switch the light if the cavity is high-loss at the DF; the regime in which it is expected to saturate is not accessible with the components used. Note that the switching takes more power to turn on than is predicted with our zero-parameter single-spatial-mode model. One possible explanation for this discrepancy is the presence of spatial mismatches between the various modes, i.e., the signal mode that is depleted by the pump, or the DF mode that is created, is not perfectly matched to the mode of the cavity. Such spatial mismatch is also a possible explanation for the $\sim$$3\%$ residual transmitted light when the model predicts a value much closer to zero. Resolving these discrepancies will be one goal of our future investigation.

In summary, we have demonstrated, for the first time, interaction-free all-optical switching with high contrast (35:1). Such switching occurs without the need for direct interaction between the control and the signal light waves. Using DF generation in a $\chi^{(2)}$-nonlinear Fabry-P\'{e}rot cavity, we have performed systematic studies of this switching mechanism in three separate operational regimes, corresponding to where the intracavity DF is resonant with the cavity,  detuned from the cavity resonance, and subjected to high intracavity loss. All three cases lead to interaction-free switching, which was observed by measuring the signal power present in both the reflected and the transmitted ports of the device. The best performance, however, was achieved when both the signal and the DF fields are in cavity resonance. All of our experimental data are in good agreement with the predictions of the theory without the need for any fitting parameter. 

Our results highlight a new approach to realizing all-optical logic operations that overcomes several fundamental constraints as well as practical difficulties associated with the existing devices. Applying this approach to nonlinear microresonators of high $Q$-factor, high-performance switching devices can be constructed that will manifest large switching contrast, low loss, low switching power, and ultralow energy dissipation. In addition, due to the ultralow in-band noise introduced by such devices, they can potentially operate on quantum-optical signals as well. For example, using a lithium-niobate microdisk with a 1-mm diameter and $Q>10^7$, whose fabrication and operation has been well demonstrated \cite{microdisk1,microdisk2}, low loss (<10$\%$) switching can be achieved with pump-pulse energy on the order of 10\,pJ. By using tailored pump pulses, together with smaller microdisks and higher $Q$, the pump-pulse energy could be further reduced to single-photon energy, leading to deterministic quantum logic gates for single-photon signals \cite{emily_theory}.

This research was supported in part by the Defense Advanced Research Projects Agency (DARPA) Zeno-based Opto-Electronics program (grant W31P4Q-09-1-0014). We acknowledge Amar Bhagwat for the construction of the crystal mount.

\bibliographystyle{apsrev}
\bibliography{zeno_switching}

\end{document}